\documentclass{aa}  

\usepackage{graphicx}
\usepackage{txfonts}

\newcommand{\kms}{km\,s$^{-1}$}

\graphicspath{ {./figures/} }

\begin{document} 

   \title{Nebular H$\alpha$ emission in SN~Ia~2016jae}

   \author{N. Elias-Rosa\inst{1,2}
          \and P. Chen\inst{3}
          \and S. Benetti\inst{1}
          \and Subo Dong\inst{3}
          \and J.~L. Prieto\inst{4,5}
          \and E. Cappellaro\inst{1}
          \and J. A. Kollmeier\inst{6}
          \and N. Morrell\inst{7}
          \and A.~L. Piro\inst{6}
          \and M. M. Phillips\inst{7}
          }

   \institute{INAF -- Osservatorio Astronomico di Padova, vicolo dell'Osservatorio 5, Padova 35122, Italy\\
              \email{nancy.elias@inaf.it}
    \and
    Institute of Space Sciences (ICE, CSIC), Campus UAB, Carrer de Can Magrans s/n, 08193 Barcelona, Spain
    \and
    Kavli Institute for Astronomy and Astrophysics, Peking University, Yi He Yuan Road 5, Hai Dian District, Beijing 100871, China
    \and
    N\'ucleo de Astronom\'{i}a de la Facultad de Ingenier\'{i}ıa y Ciencias, Universidad Diego Portales, Av. Ej\'eercito 441, Santiago, Chile
    \and
    Millennium Institute of Astrophysics, Santiago, Chile
    \and
    Observatories of the Carnegie Institution for Science, 813 Santa Barbara Street, Pasadena, CA 91101, USA
    \and
    Las Campanas Observatory, Carnegie Observatories, Casilla 601, La Serena, Chile 
    }

   \date{Received XXXX; accepted XXXX}

 
  \abstract
   {There is a wide consensus that type Ia supernovae (SN Ia) originate from the thermonuclear explosion of CO white dwarfs (WD), with the lack of hydrogen in the observed spectra as a distinctive feature. Here, we present SN~2016jae, which was classified as a Type Ia SN from a spectrum obtained soon after the discovery. The SN reached a B-band peak of -17.93 $\pm$ 0.34 mag, followed by a fast luminosity decline with s$_{BV}$ 0.56 $\pm$ 0.06 and inferred $\Delta$m$_{15}(B)$ of 1.88 $\pm$ 0.10 mag. Overall, the SN appears as a ``transitional'' event between ``normal'' SNe~Ia and very dim SNe Ia such as 91bg-like SNe. Its peculiarity is that two late-time spectra taken at +84 and +142 days after the peak show a narrow line of H$\alpha$ (with full width at half-maximum of $\sim$ 650 and 1000 km\,s$^{-1}$, respectively). This is the third low-luminosity and fast-declining Type Ia SN after SN2018cqj/ATLAS18qtd and SN2018fhw/ASASSN-18tb, found in the 100IAS survey that shows resolved narrow H$\alpha$ line in emission in their nebular-phase spectra. We argue that the nebular H$\alpha$ emission originates in an expanding hydrogen-rich shell (with velocity $\le$ 1000 km\,s$^{-1}$). The hydrogen shell velocity is too high to be produced during a common envelope phase, while it may be consistent with some material stripped from an H-rich companion star in a single-degenerate progenitor system. However, the derived mass of this stripped hydrogen is $\sim$0.002-0.003 M$_{\sun}$, which is much less than that expected ($>$0.1 M$_{\sun}$) for standard models for these scenarios. Another plausible sequence of events is a weak SN ejecta interaction with a H-shell ejected by optically thick winds or a nova-like eruption on the C/O WD progenitor some years before the supernova explosion.
    }

   \keywords{supernovae: general;
             supernovae: individual: SN~2016jae;
             supernovae: individual: SN~2018cqj;
             supernovae: individual: SN~2018fhw
               }

   \maketitle

%
\section{Introduction}\label{intro}

Type Ia supernovae (SN~Ia) are believed to originate from the thermonuclear explosion of CO white dwarfs (WD). In a popular explosion scenario, the WD is in a close binary system with another WD (double degenerate system), or with a non-degenerate mass-donor companion, either a main sequence, or a giant star (single degenerate system). In the merging scenario, the combined mass of the WDs may meet or exceed the Chandrasekhar limit ($\sim$ 1.4 M$_{\sun}$), while in the single degenerate, the mass transfer from the secondary star makes the primary WD approaches the Chandrasekhar limit. In both cases, the final fate of the system is the explosion. There are also alternative scenarios in which the explosion occurs without reaching the Chandrasekhar mass (i.e., a sub-Chandrasekhar-mass explosion). Which is the dominant scenario is still unknown (see, e.g., \citealt{maoz14}). 
A distinctive feature of SNe Ia is the lack of hydrogen in the observed spectra. Yet, at least for single degenerate systems, it is expected that some H from the main-sequence companion star is left in the system. In fact, finding H in SNe~Ia would strongly support the single degenerate scenario. Several attempts have been made to detect H in SNe Ia, but no convincing evidence has been found yet (e.g. \citealt{mattila05,tucker20}). 
Hydrogen has been found so far only in the SNe Ia-CSM (circumstellar material) events, where narrow H emissions are present since early phases and probably excited in the shock between the SN ejecta and H-rich CSM (e.g. \citealt{hamuy03,prieto07,dilday12,silverman13}; consider also \citealt{benetti06} for an alternative interpretation). 

Recently, two cases have been reported of low-luminosity SNe~Ia with detection of narrow H$\alpha$ emission lines in late-time spectra: SN~2018fhw/ASASSN-18tb \citep{kollmeier19,vallely19} and SN~2018cqj/ATLAS18qtd \citep{prieto20}. These two SNe were followed as part of the 100IAS survey \citep{dong18}, a project aiming at collecting a homogeneous sample of $\sim$ 100 SNe~Ia with optical nebular-phase spectra using 5-10 meter class telescopes. Both SNe show strong emission lines of H$\alpha$ at phases $>$ 100 d, with luminosities in the range of $\sim$ 10$^{36}$ - 10$^{38}$ erg\,s$^{-1}$ (H$\alpha$ emission lines in SNe~Ia-CSM can reach luminosities in the range 10$^{40}$ - 10$^{41}$ erg\,s$^{-1}$; \citealt{silverman13}). 

Here we report a third case, SN~2016jae ($\alpha$ = 09$^{\rm h}$ 42$^{\rm m}$ 34${\fs}$49, $\delta$ = +10$^{\circ}$ 59$\arcmin$ 35${\farcs}$7; J2000.0), also known as ATLAS16eay, MASTER OT J094234.49+105935.7, CSS170201-094234+105935, Gaia16cev and PS16fkf. It was discovered on 2016 December 21.99 UT, with an unfiltered magnitude of 17.2, by the MASTER Global Robotic Net \citep{atel9902}\footnote{\url{http://master.sai.msu.ru/en/}}. Independent discoveries were also reported by the Asteroid Terrestrial-impact Last Alert System (ATLAS, \citealt{tonry18}, \citealt{smith20}; \citealt{tns1077}), Gaia transient survey \citep{hodgkin13} and Pan-STARRS (\citealt{chambers16}, \citealt{magnier16}).
It was classified on 2016 December 28.34 UT \citep{atel9908,tns1095} as a Type Ia SN approximately one week post-maximum, by the Public ESO -European Southern Observatory- Spectroscopic Survey for Transient Objects (PESSTO; \citealt{smartt15}).

In the next section (Section \ref{SNhostgx}), we describe the host environment of SN~2016jae. Photometric and spectroscopic data are analysed in Section \ref{sn16jae}. And the discussion is in Section \ref{nature16jae}. 
%
\section{Host environment}\label{SNhostgx}

\begin{figure}[!ht]
\centering
\includegraphics[width=0.9\columnwidth]{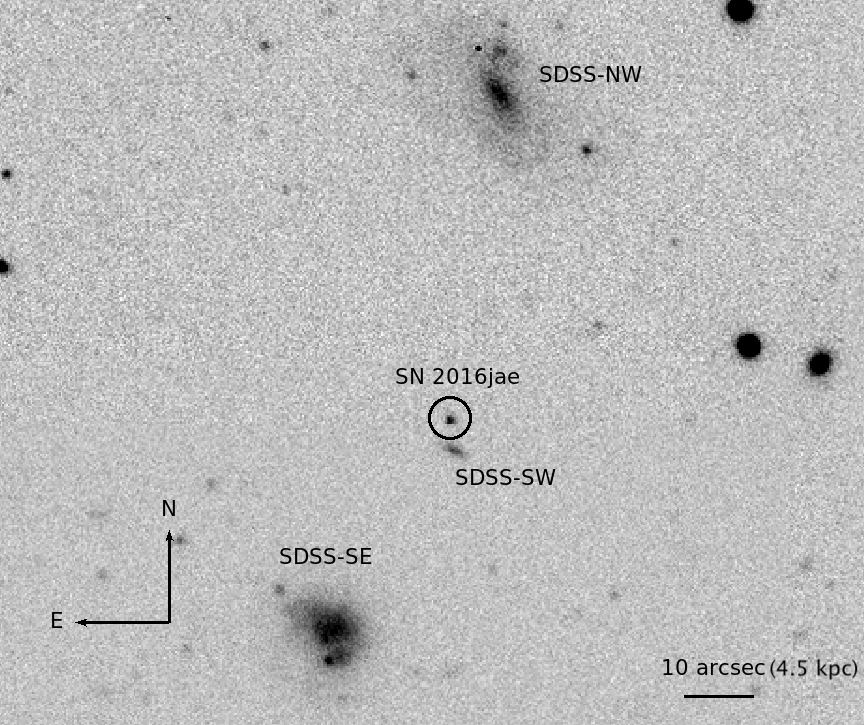}
\caption{DECals DR7 $r$ deep image of the SN~2016jae field. The SN and the three nearby galaxies discussed in the text (SDSS-SW, SDSS-SE, and SDSS-NW) are indicated. We consider D$_L$ = 92.9 $\pm$ 4.3 Mpc to estimate the physical projected distance in the scale bar. We argue that SN~2016jae exploded most likely in the outer halo of SDSS-SE, as this is the nearest galaxy with a similar redshift.}
\label{fig_FC}%
\end{figure}

The galaxy in the closest angular separation with SN~2016jae is SDSS~J094234.46+105931.1 (hereafter SDSS-SW), a tiny galaxy located at 4${\farcs}$2 S, 0${\farcs}$6 W from the SN (see Fig. \ref{fig_FC}). The candidate host galaxy has a magnitude in r of 21.67 and a photometric redshift of 0.35 $\pm$ 0.08 according to the Sloan Digital Sky Survey (SDSS) DR15 catalogue\footnote{\url{http://skyserver.sdss.org/dr15/en/home.aspx}}. However, the redshift derived from the spectrum of SN~2016jae using the Supernova Identification cross-correlation code (SNID; \citealt{blondin07}), z = 0.021 $\pm$ 0.006, is not consistent with the redshift of that galaxy. 

There are other two galaxies in the proximity of SN~2016jae, SDSS~J094235.65+105905.3 (hereafter SDSS-SE), and SDSS~J094234.01+110022.8 (hereafter SDSS-NW) located at 30${\farcs}$0 S, 17${\farcs}$3 E and 47${\farcs}$5 N, 7${\farcs}$3 W, respectively, from the SN. We took a spectrum of SDSS-SW and SDSS-SE with the 10~m Gran Telescopio Canarias (GTC) on 2019 December 3, and another spectrum of SDSS-SW and SDSS-NW always with GTC on 2020 May 10. Both spectra were reduced as described in Section \ref{SNspec}. In the case of SDSS-SW, we combine the spectra of different epochs to improve the signal-to-noise ratio (see Fig. \ref{fig_spechost}). Comparing the spectrum of SDSS-SW with a template spectrum of a late-type galaxy (from the Kinney-Calzetti Spectral Atlas; \citealt{calzetti94,kinney96}) and measuring the position of H$\alpha$, H$\beta$, [\ion{O}{ii}] and [\ion{O}{iii}] emission lines, we found that SDSS-SW is consistent with being a background galaxy at z=0.285 $\pm$ 0.009 (similarly to the SDSS photometric redshift). Measuring the position of the same emission lines, plus [\ion{N}{ii}] and [\ion{S}{ii}] in the SDSS-SE spectrum, and H$\alpha$, [\ion{O}{iii}] and [\ion{S}{ii}] in SDSS-NW, we derive a redshift of 0.021 $\pm$ 0.001 and 0.019 $\pm$ 0.001, respectively, which are consistent with the redshift of the SN derived from fitting its near-peak spectrum. This suggests that SN~2016jae exploded in the outer halo of one of these galaxies, most likely in SDSS-SE for being the closest one. Throughout the paper, we adopt z$_{\rm host}$ = 0.021 of SDSS-SE as the reference redshift. However, we cannot rule out that the SN is actually located in the halo of SDSS-NW. The implication assuming z$_{\rm host}$ = 0.019 is discussed in Sect. \ref{nature16jae}. From the adopted redshift we derived a luminosity distance of 92.9 $\pm$ 4.3 Mpc\footnote{An uncertainty of the velocity of 250 km\,s$^{-1}$ was added in quadrature to the distance error (see, \citealt{burns18}).} (m-M = 34.8 $\pm$ 0.1 mag), assuming a Hubble constant H$_0$ = 67.8 km\,s$^{-1}$\,Mpc, $\Omega_{m}$ = 0.31 and $\Omega_{\Lambda}$ = 0.69 \citep{planck16}. With this distance, the projected linear offset from the SDSS-SE nucleus is 15.6 kpc.

The Milky Way extinction along the line of sight of the supernova is $A_{V, \rm MW} = 0.051$ mag (NED\footnote{The NASA/IPAC Extragalactic Database (NED) is funded by the National Aeronautics and Space Administration and operated by the California Institute of Technology.}; \citealt{schlafly11}). 
There is no evidence of narrow \ion{Na}{i}D absorption in the spectra, which would indicate the presence of gas/dust on the line of sight. Taking advantage of the fact that SNe Ia are known for their ``homogeneity'', we considered different methods to estimate the extinction (A$_V$) toward SN~2016jae based on comparisons of the object’s spectral energy distribution (SED) and luminosity with those of other similar SNe~Ia. As we will discuss in the next section, SN~2016jae is a transitional event between ``normal'' SNe~Ia and sub-luminous SNe~Ia such as SN~1991bg. Then, we matched the intrinsic (B$-$V)$_0$, and (r$-$i)$_0$ colour curves of SN~2016jae with those of SN~2018cqj \citep{prieto20}, SN~2018fhw \citep{vallely19} (which have a similar decline rate), and SN~1991bg, the prototype of sub-luminous SNe Ia \citep{filippenko92,leibundgut93,turatto96}, obtaining E(B-V)$_{V,tot}$ = 0.09 $\pm$ 0.03 mag (see Figure \ref{fig_colourevol}). For comparison we also show the colour curves of the ``normal'' Type~Ia SNe~1992A \citep{kirshner93} and 2011fe \citep{pereira13}. Besides, we contrasted the early-time optical SED of SN~2016jae with those of 91bg-like SNe~Ia at similar epoch. The SED of the reference SNe were first corrected for redshift and Galactic extinction. From this comparison a colour excess of E(B-V)$_{V,host}$ = 0.09 $\pm$ 0.01 mag was derived. We adopt E(B-V) = 0.10 $\pm$ 0.03 mag, i.e., A$_V$ = 0.30 $\pm$ 0.10 mag, as the total extinction toward SN~2016jae.

\section{Data and analysis}\label{sn16jae}

\subsection{Photometry}\label{SNph}

We used $BV ri$ photometry of SN~2016jae between 2016 December 28 (close to its peak magnitude) and 2017 February 13 from the CNIa0.02 programme\footnote{Collection of a Complete, Nearby, and effectively unbiased sample of Type Ia Supernovae at host-galaxy redshifts z$_{host} <$ 0.02.} \citep{chen20} and taken with the 1~m telescopes at the Las Cumbres Observatory Global Telescope Network (LCOGT; \citealt{brown13}).
We also retrieved three epochs of $gaia$-band photometry from the Gaia transient survey\footnote{\url{http://gsaweb.ast.cam.ac.uk/alerts}}, and a late-time (phase 142d) $r$-band photometric measurement of SN~2016jae with the 10.4~m GTC+OSIRIS at the Roque de los Muchachos Observatory (Spain). The Gaia data were converted to $r$-band magnitude using the relationships reported in Gaia Data Release 2 Documentation (release 1.2)\footnote{\url{https://gea.esac.esa.int/archive/documentation/GDR2/Data_processing/chap_cu5pho/sec_cu5pho_calibr/ssec_cu5pho_PhotTransf.html}}. GTC SN magnitudes were measured using Point-Spread-Function (PSF) fitting with SNOoPy\footnote{SNOoPy is a package for SN photometry using PSF fitting and/or template subtraction developed by E. Cappellaro. A package description can be found at \url{http://sngroup.oapd.inaf.it/ecsnoopy.html}.} pipeline. We calibrated the instrumental magnitudes to standard photometric systems, using the zero points and colour terms measured through reference SDSS stars in the field of SN~2016jae.
We present the final calibrated photometry of SN~2016jae in Table \ref{table_ph}.\\

\begin{figure*}[!ht]
\centering
\includegraphics[width=0.49\textwidth]{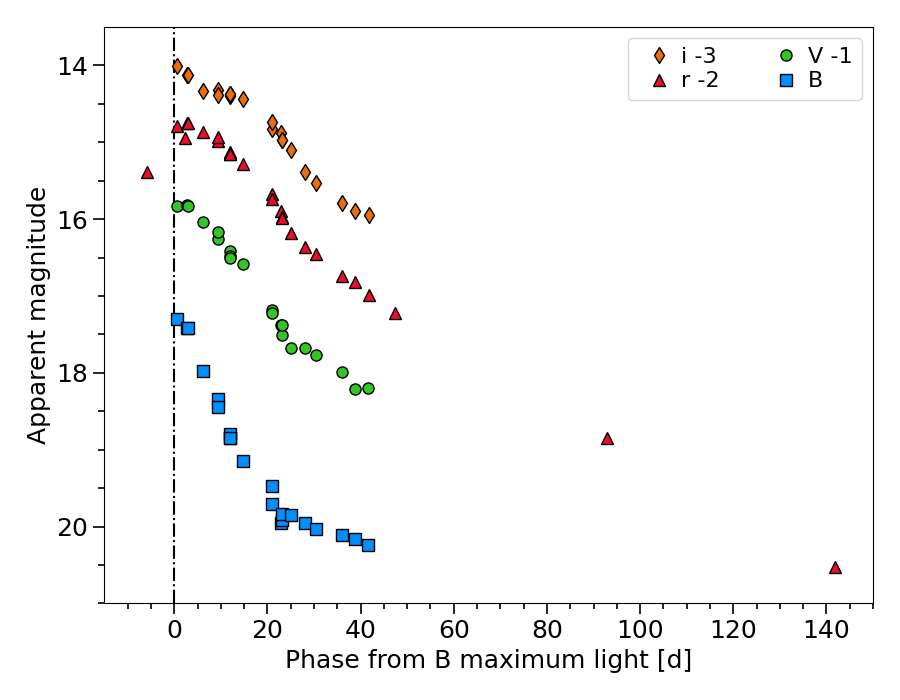} \hfill 
\includegraphics[width=0.49\textwidth]{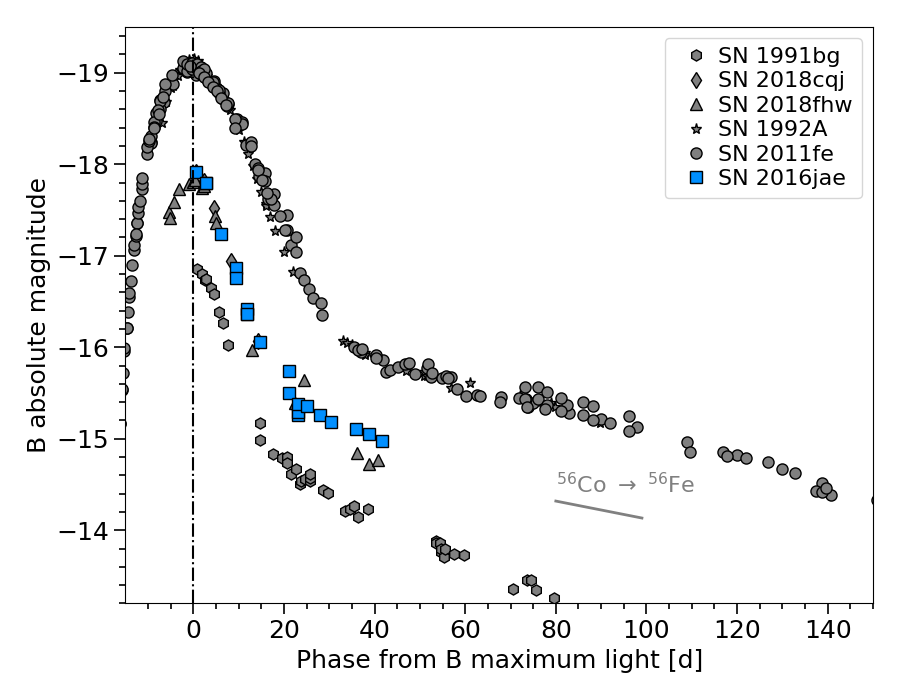}
\caption{{\it Left:} $BV ri$ light curves of SN~2016jae. The light curves have been shifted for clarity by the amounts indicated in the legend. {\it Right:} Absolute $B$ light curve of SN~2016jae, shown along with those of SNe~1991bg, 2018cqj, 2018fhw, 1992A and 2011fe. For both panels, the dot-dashed vertical line indicates the $B$-band maximum light. The uncertainties for most data points are smaller than the plotted symbols. A colour version of this figure can be found in the online journal.}
\label{fig_lc}%
\end{figure*}

The $BV ri$ light curves of SN~2016jae are shown in the left panel of Figure \ref{fig_lc}. The data are relative to the $B$ maximum date occurring on 2016 December 28.2, or MJD 57750.2 $\pm$ 1.0. It was obtained by matching the light curves peak with the {\sc SNooPy} fitting package \citep{burns11} using the colour stretch parameter s$_{BV}$ models. Allowing for the poor sampling around the maximum light, we obtained the following fits: B$_{max}$ = 17.32 $\pm$ 0.11 mag, V$_{max}$ = 16.77 $\pm$ 0.07 mag, r$_{max}$ = 16.78 $\pm$ 0.07 mag, i$_{max}$ = 17.07 $\pm$ 0.08 mag and s$_{BV}$ = 0.56 $\pm$ 0.06. 
The light curves of SN~2016jae have a good sampling after maximum light up to +42 d later, showing in all bands a fast decline until the radioactive decay at t $\sim$ 20 d in $B$-band. The latter translates to $\Delta$m$_{15}$($B$) of 1.88 $\pm$ 0.10 mag using the s$_{BV}$-$\Delta$m$_{15}$ relationship in \cite{burns18}. 
Based on the light curves, SN~2016jae qualifies as an intermediate case between ``normal'' and sub-luminous and fast-declining Type Ia supernovae. In fact,

\begin{itemize}
\item The $\Delta$m$_{15}$ of SN~2016jae is between that typical for ``normal'' SNe~Ia, $\lesssim$ 1.7 mag, and that of 91bg-like SNe, 1.8-2.1 mag (see, e.g., \citealt{taubenberger17}).

\item The $B$ absolute magnitude peak of SN~2016jae is -17.93 $\pm$ 0.34, dimmer than ``normal'' Type Ia SNe such as SNe~1992A and 2011fe (see Figure \ref{fig_lc}, right panel) and similar to other fast-decliner objects such as SNe~2018cqj and 2018fhw. 
We note that the latter SNe are more luminous than SN~1991bg, as illustrated in Figure \ref{fig_lc}. Indeed, the location of SN~2016jae in the figure M$_{B_{max}}$ vs. (B$_{max}$-V$_{max}$) colour (figure 7 of \citealt{phillips12}) falls rather close to transitional supernovae such as the 86G-like (see Fig. \ref{fig_BVvsMB}).

\item The ``pseudo-bolometric'' (integrating the flux in the $BV ri$ bands) peak luminosity of SN~2016jae, log(L) = 42.49 $\pm$ 0.07 is consistent with those of other sub-luminous Type Ia supernovae.

\item Differently from 91bg-like SNe, the light curve in $i$-band shows a flattening at $\sim$ 20 days, yet not a secondary peak as typically seen in the redder light curves of ``normal'' SNe Ia. According to \citealt{kasen06} and \citealt{blondin15}, the secondary near-infrared peak in SNe~Ia is related to a recombination wave propagating through chemically stratified ejecta. Consequently, the recombination of \ion{Fe}{iii} to \ion{Fe}{ii} would occur earlier in less luminous and cooler SNe. 

\item We can estimate the mass of $^{56}$Ni synthesized in an explosion through the bolometric peak \citep{stritzinger05}. We derived a lower limit (because we have integrated only the flux in the $BV ri$ bands) for the $^{56}$Ni mass of 0.15 M$_{\sun}$ for SN~2016jae. Similar masses have been also estimated for SNe~2018cqj and 2018fhw\footnote{Our estimation of $^{56}$Ni mass for SN~2018fhw is consistent with the value derived by \citet{vallely19}, 0.2 M$_{\sun}$} (computed in a similar manner). This is marginally larger than the range found for the 91bg-like SNe (between 0.05 and 0.10 M$_{\sun}$; \citealt{taubenberger17}).

\end{itemize}

\subsection{Spectroscopy}\label{SNspec}

We analyze three optical spectra of SN~2016jae (see Table \ref{table_spec} for basic information on the spectroscopic observations). The first spectrum was taken for the transient classification at the 3.6m New Technology Telescope (NTT) on 2016 December 28.34 by the international collaboration PESSTO\footnote{The spectrum was taken from WiseREP \citep{yaron12}; \url{https://wiserep.weizmann.ac.il}.}. Two more spectra were obtained at nebular phase, on 2017 March 22.19 and 2017 May 18.91, with the Magellan Clay telescope (as part of 100IAS) at Las Campanas Observatory and with GTC at Roque de los Muchachos Observatory, respectively. All the spectra were obtained with the slit aligned with the parallactic angle to minimize differential flux losses caused by atmospheric refraction.

All spectra were reduced following standard procedures with {\sc IRAF} routines. The two-dimensional frames were corrected by bias and flat-field, before the extraction of the one-dimensional spectra. 
We wavelength calibrate the spectra by comparison with arc-lamp spectra. The flux calibration was done using spectrophotometric standard stars, which also help for removing the strongest telluric absorption bands present in the spectra (in some cases, residuals are still present after the correction). The wavelength calibration was verified against the bright night-sky emission lines. Finally, the absolute flux calibration of the spectra was cross-checked against the broadband photometry: $BV ri$ SED at a contemporaneous phase for the first spectrum, $V ri$ extrapolated SED for the second spectrum and $r$ band for the late time spectrum. In all three cases, we scaled the spectra by a constant value (ranging from 0.8 to 1.9), assuring a final accurate match with the photometry of $<$ 0.15 mag.

\begin{figure}[!ht]
\centering
\includegraphics[width=\columnwidth]{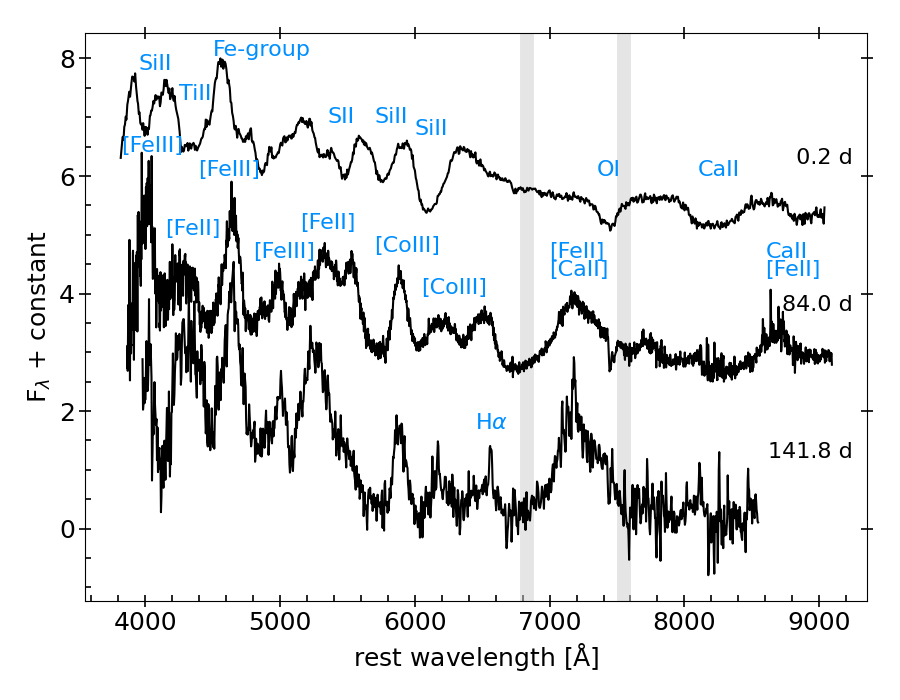} 
\caption{Optical spectral sequence of SN~2016jae. All spectra have been corrected by redshift and extinction. The grey columns show the location of the strongest telluric band, which has been removed when possible. The locations of the most prominent spectral features are also indicated. A colour version of this figure can be found in the online journal.}
\label{fig_specevol}%
\end{figure}

Figure \ref{fig_specevol} shows the three optical spectra of SN~2016jae. The first spectrum of SN~2016jae is similar to that of ``normal'' Type Ia SNe, except for the unusually prominent \ion{O}{i} $\lambda$7774 and \ion{Ti}{ii} (between 4000 and 4400 \text{\AA}) features, more common of faster-decliner and subluminous SNe~Ia such as 91bg-like SNe and transitional events like SN~1986G (\citealt{taubenberger17}; see also panel (a) of Figure \ref{fig_spececomp_earlylate}). In fact, $\cal R$(\ion{Si}{ii}) \citep{nugent95}, the ratio between the strength of the absorptions of \ion{Si}{ii} $\lambda$5972 and $\lambda$6355, is close to that of the FAINT and CL ('cool') families according to \citet{benetti05} and \citet{branch06}, respectively\footnote{FAINT SNe Ia (one of the three families grouped by \citealt{benetti05} according they observed diversities) are characterized by having low expansion velocities and rapid evolution of the \ion{Si}{ii} velocity. In the classiﬁcation scheme of \citet{branch06}, the CL or cool SNe Ia are those with moderate values of pseudo-equivalent width of the \ion{Si}{ii} $\lambda$6355 line, but large values of pseudo-equivalent width of the \ion{Si}{ii} $\lambda$5972 line.}. The late-time spectra of SN~2016jae also exhibit features previously seen in 91bg-like and transitional SNe~Ia. The spectra are dominated by narrow [\ion{Fe}{ii}], [\ion{Fe}{iii}] and [\ion{Co}{iii}], and broad [\ion{Ca}{ii}]/[\ion{Fe}{ii}] emission lines (panel (b) of Figure \ref{fig_spececomp_earlylate}).

\begin{figure}[!ht]
\centering
\includegraphics[width=0.9\columnwidth]{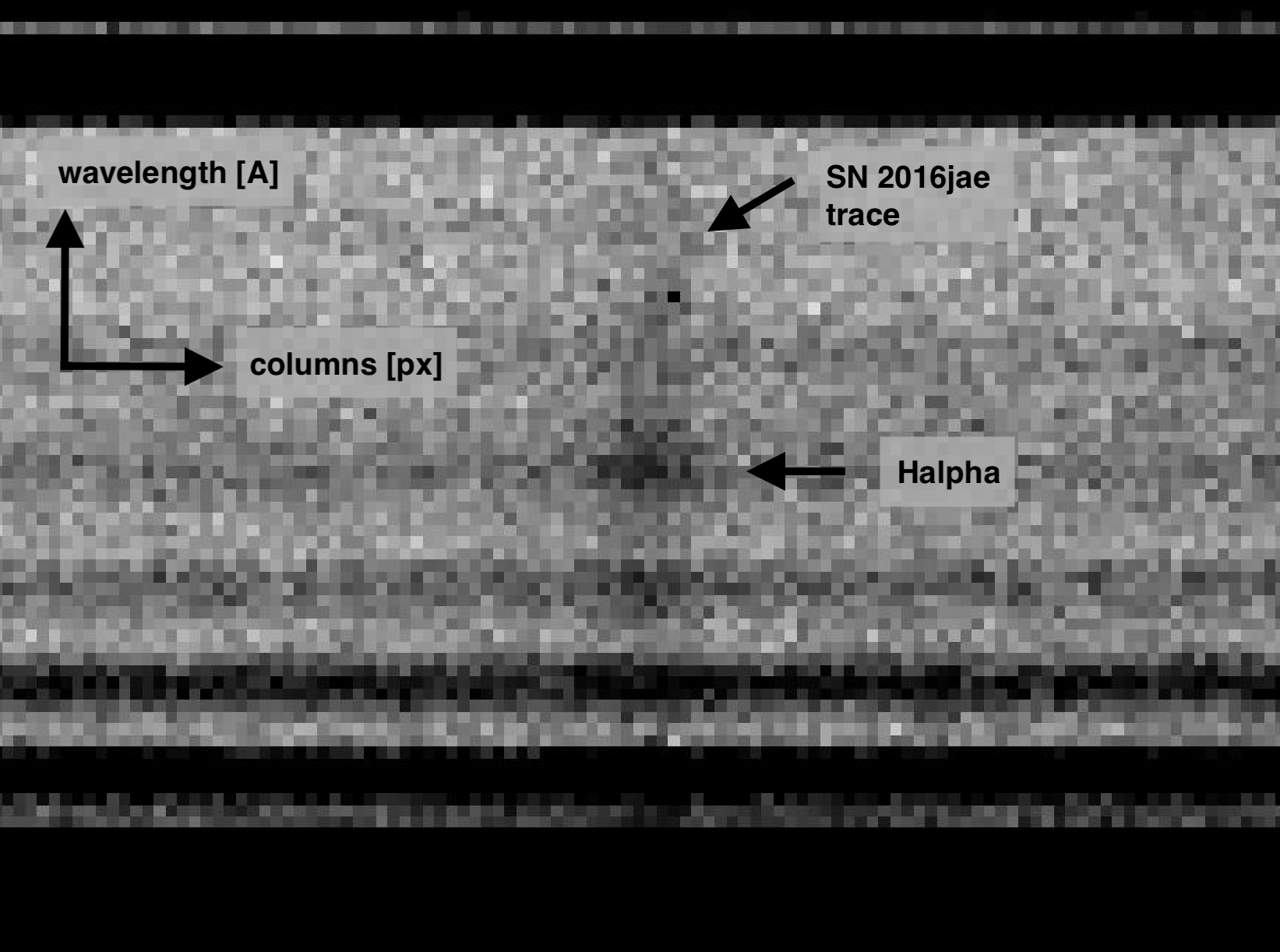} 
\caption{The H$\alpha$ emission line in the 2D spectrum of SN~2016jae taken with GTC+OSIRIS on 2017 May 18 (phase 141.8 d from the assumed B -maximum date). The wavelength is up and the spatial direction is to the right.}
\label{fig_halpha2D}%
\end{figure}

\begin{figure*}[!ht]
\centering
\includegraphics[width=0.8\textwidth]{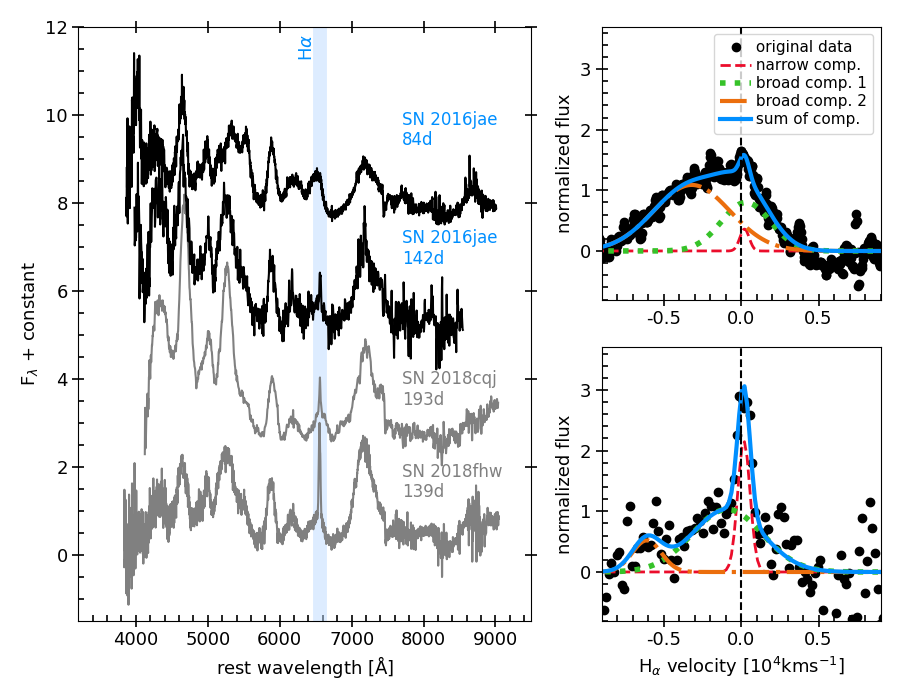} 
\caption{{\it Left:} Comparison of SN~2016jae late-time optical spectra, along with those of the SNe~2018cqj and 2018fhw at similar epochs. All spectra have been corrected by redshift and extinction (values adopted from the literature; see also Table \ref{table_SNe}). Ages are relative to $B$ maximum light.  
The locations of the H$\alpha$ emission line is also indicated. {\it Right:} Decomposition of the H$\alpha$ emission line of SN~2016jae at phases 84.0 ({\it top}) and 141.8 d ({\it bottom}). Three Gaussian profiles have been used for the narrow H$\alpha$ and two broad components of iron-group elements.}
\label{fig_spececomp_late}%
\end{figure*}

Motivated by the two preceding discoveries of nebular H$\alpha$ lines in 100IAS, we inspected the spectra of SN~2016jae and noticed the presence of a significant H$\alpha$ emission line. It is weak and barely visible at phase 84.0d, but well seen at phase 141.8d (see also the 2D image of the spectrum in Figure \ref{fig_halpha2D} and the H$\alpha$ line profile comparison in Figure \ref{fig_halphacomp_SNhost}, which support our belief that the H$\alpha$ line is intrinsic to the SN spectrum and not a contamination from nearby galaxy background).
This same line is also present in the nebular spectra of SNe~2018cqj and 2018fhw at similar phases, as can be seen in Figure \ref{fig_spececomp_late}, left panel. 
The H$\alpha$ line of SN~2016jae appears on top of a broader feature, probably due to a blend of iron-group elements (see, e.g., \citealt{mazzali97}). We decompose the line profiles into three Gaussian profiles, one for the narrow H$\alpha$ and two components to match the broad iron-peak element feature (see \citealt{eliasorsa16} for more details in the procedure). The right panels of Figure \ref{fig_spececomp_late} show the result of the multi-component fit. The profiles are well reproduced with a narrow component centered at $\sim$ 6568 \text{\AA} rest frame, with full width at half-maximum (FWHM) of $\sim$ 16 \text{\AA} ($\sim$ 650 km\,s$^{-1}$; after correction for instrumental resolution\footnote{For the resolved narrow line components, we first corrected the measured FWHM for the spectral resolution ($width = \sqrt{FWHM^{2} - res^{2}}$) and then computed the velocity ($v = width \times c$).}) at phase 84.0d, and at $\sim$ 6567 \text{\AA}, with FWHM of $\sim$ 22 \text{\AA} ($\sim$ 1000 km\,s$^{-1}$) at phase 141.8d. The integrated luminosity of H$\alpha$ was estimated to be L(84.0d) = (3.0 $\pm$ 0.8 $\times$ 10$^{38}$ erg\,s$^{-1}$, and L(141.8d) = (1.6 $\pm$ 0.2) $\times$ 10$^{38}$ erg\,s$^{-1}$. The best-fit luminosity of the H$\alpha$ line decreases by a factor of $\sim$ 2 between the 84.0 and 141.8 days spectra, however, given the error of the flux measurement in the first epoch we cannot exclude that the flux could remain constant.
In the case of SN~2016jae, as well as in SNe 2018cqj and 2018fhw, the H$\beta$ is not visible. Following the method introduced by \citet{tucker20}, we have estimated a 10$\sigma$ upper limit on the H$\beta$ flux of the order of $\sim$ 10$^{-17}$ erg\,s$^{-1}$\,cm$^{-2}$. A similar result was also obtained for SNe 2018cqj and 2018fhw. 

%
\section{Discussion and summary on the nature of SN~2016jae and its progenitor scenario}\label{nature16jae}

In the previous sections, we have analyzed the observed properties of SN~2016jae. This supernova, classified as a Type Ia, shows a B-band peak of -17.93 $\pm$ 0.11 mag, an intermediate value between that of SN~2011fe (``normal'' SN; M$_{B,max}$ = -19.1, \citealt{ashall16}) and SN~1991bg (sub-luminous SN prototype; $>$ -17.7 mag, \citealt{taubenberger17}), followed by a fast decrease with $\Delta$m$_{15}$ of 1.88 $\pm$ 0.10 mag. 
Despite its early classification as a ``normal'' SN Ia, the SN observables of SN~2016jae point to a similarity to fast declining and sub-luminous transitional events. The noticeable feature is a narrow H$\alpha$ line that appears in the nebular spectra. This emission line is associated with the SN, and the velocities are 650 and 1000 km\,s$^{-1}$ at phase 84d and 142d, respectively. The derived FWHM is similar to those obtained for SNe~2018cqj and 2018fhw ($\sim$ 1200 km\,s$^{-1}$ at 193d and 139d, respectively), the other two low-luminosity SNe~Ia with detection of H$\alpha$ emission line in late-time spectra.

Understanding the origin of this H emission can give crucial constraints for the progenitor scenario. We first consider two scenarios: 
(i) pre-existing CSM material shocked by the SN ejecta, or (ii) in the material stripped by the SN ejecta from an H-rich companion star in a single-degenerate progenitor system. 

Nearby CSM can originate from the system before explosion due to the outcome of a common envelope phase. In this case, the expected expansion velocity of H$\alpha$ is of the order of a few hundred km\,s$^{-1}$, comparable to that presumed for interacting SNe (e.g. \citealt{silverman13,smith14}). However, the estimated velocities for SNe~2016jae, 2018cqj, and 2018fhw are much higher.
Besides, H$\alpha$ should also appear at early phases and accompanied by other Balmer emissions as H$\beta$ (we noticed the non-detection of H$\beta$ in SN~2016jae in Section \ref{SNspec}). 
In short, it is unlikely that the gas shell was produced by a common envelope phase.
Also, the parent stellar population of these three transitional SNe Ia with H$\alpha$ emission was most likely of old age. In fact, both SNe~2018cqj and 2016jae appeared in the outskirts of their putative host galaxies, suggesting either that they are halo objects, whereas the host of SN~2018fhw was a dwarf elliptical. This is consistent with the progenitor environment of the majority of the 91bg-like and transitional SNe (see, e.g., \citealt{panther19}), rather than with that of Type Ia-CSM SNe, which are found in star-forming galaxies (see, e.g., \citealt{hamuy03,prieto07,dilday12,silverman13}).

An alternative is that SN~2016jae exploded in a single degenerate system, and the H material was stripped from the non-degenerate H-rich companion star.
It is expected that the stripped material has expansion velocities $\leq$ 1000 km\,s$^{-1}$ (see, e.g., \citealt{mattila05,liu12}) that is consistent with the estimated velocity from the H$\alpha$ line in SN~2016jae. 
In the recent simulations, \citet{dessart20} found that the hydrogen stripped from a non-degenerate companion star can remain undetected at epochs $<$50d, begin to emerge at 100d and well visible between 150d and 200d (at this time the metal-rich ejecta become increasingly more transparent and faint). This is fully consistent with our observations of SN~2016jae. Besides, these authors argue that we may expect to see H$\alpha$ and not H$\beta$ because H$\alpha$ sits in a most transparent regions of the optical range, while H$\beta$ occurs in a region with strong metal line blanketing. 
Notice that stripping could also be asymmetrical and the cause of the possible redshift of $\sim$ 200 km\,s$^{-1}$ in the H$\alpha$ profile (at both 84d and 142d phases for the adopted z$_{host}$ = 0.021; \citealt{prieto20,botyanszki18}). Actually, \citet{prieto20} stressed that the apparent line shift may be related to the uncertainties in the location and redshift of the SN. It may be a concern for SN~2016jae because as discussed in Section \ref{SNhostgx}, it is not clear which galaxy hosts this SN. However, assuming z$_{\rm host}$ = 0.019, the H$\alpha$ peak would be even more blue-shifted by $\sim$ 800 km\,s$^{-1}$.

\begin{figure}[!ht]
\centering
\includegraphics[width=0.9\columnwidth]{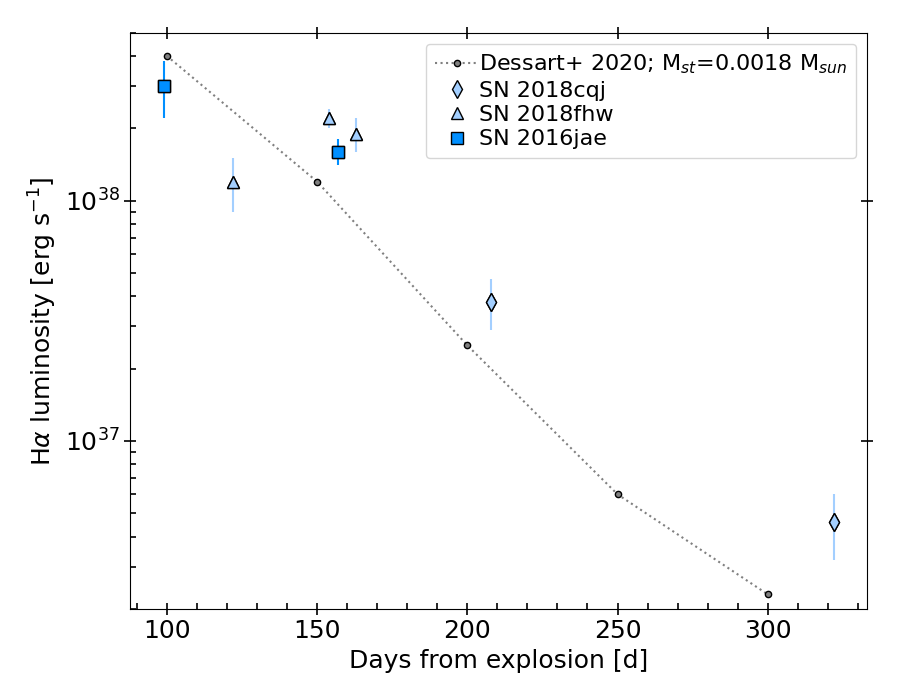} 
\caption{Evolution of the luminosity of H$\alpha$ of SNe~2016jae, 2018cqj, 2018fhw. For comparison, we have also plotted an average value for the subset of models with M$_{st}$ = 0.0018 M$_{\sun}$ of \citet{dessart20} (dotted line). For the SNe of the sample, we  have assumed 15 days of rise-time (average from the values found by \citealt{hsiao15} and \citealt{vallely19}).}
\label{fig_lumhalpha}%
\end{figure}

As in SNe~2018cqj and 2018fhw, we can use the luminosity estimation from the nebular spectra and constrain the amount of hydrogen stripped from the non-degenerated companion star. Using the relation introduced by \citet{tucker20}, which is based on the models of \citet{botyanszki18}, we can derive a mass of the hydrogen stripped material (M$_{st}$) of 0.002 M$_{\sun}$ from the two nebular spectra of SN~2016jae. Similarly, following \citet{dessart20} and considering the estimated $^{56}$Ni mass for SN~2016jae in Section \ref{SNph} (0.15 M$_{\sun}$\footnote{The value of $^{56}$Ni mass allows us to discriminate between the models used by \citet{dessart20} for the analytical fit to the correlation between the luminosity of H$\alpha$ and M$_{st}$. Here we have followed the delayed-detonation model DDC25 with $^{56}$Ni mass of 0.12 M$_{\sun}$.}), we obtain a comparable value for M$_{st}$ of 0.003 M$_{\sun}$. This estimated M$_{st}$, is similar to that found for SNe~2018cqj and 2018fhw, and as already notice by \citet{dessart20}, it is much lower than expected ($>$0.1 M$_{\sun}$) for standard models of SD scenarios for SNe Ia (see, e.g., \citealt{marietta00,liu12,botyanszki18}). 
We attempted to understand the evolution with time of the H$\alpha$ luminosity considering the measurements for the three SNe with hydrogen detected. As we see in Figure \ref{fig_lumhalpha}, the H$\alpha$ luminosity decreases with the phase of the SN. 
\citet{dessart20} showed that the luminosity of H$\alpha$ accounts for 10$\%$ of the radioactive decay energy absorbed by the stripped gas.
This dependence implies a connection between the H$\alpha$ luminosity and the characteristic time of the $^{56}$Co decay. That is, the luminosity of H$\alpha$ decreases with time, and this is faster when M$_{st}$ is low. The latter is related to the fact that for low M$_{st}$, $\gamma$ rays are not efficiently trapped by the stripped material.  

To check our estimates of the stripped mass derived from the H$\alpha$ luminosity, we have measured the equivalent width (EQW) of this emission line, which is a parameter that does not depend on the distance or extinction of the object. For this, we have followed the steps described in \citet{dessart20} and applied their analytical fit for a DDC25 model. We obtained a M$_{st}$ of 0.014 and 0.066 M$_{\sun}$ at 84d and 142d, respectively, that are consistent with the values derived above. 

In conclusion, while the H emission observed in SN~2016jae may be from material stripped from a companion, it seems in conflict with the M$_{st}$ predicted by hydrodynamic models for this type of scenario. Yet, we may consider some mechanism to hide the stripped hydrogen gas. For example, some of the H-rich material may be not visible because it is travelling at higher velocities (although models predict that the amount of this high-velocity material is low; see, e.g., \citealt{pan12,liu12}). Otherwise, the outer layer of the companion star might have been stripped off by an optically dense wind well before the explosion \citep{hachisu99,hachisu08}.\\

A key characteristic of SN~2016jae is the H$\alpha$ expansion velocity of $\sim$ 1000 km\,s$^{-1}$. Such velocity can be reached by optically thick winds and nova ejecta (\citealt{hachisu99,moriya19}).

Based on this suggestion, one may consider a binary system made of a WD and a secondary H-rich star which is transferring mass to the WD. When the mass transfer rate becomes larger than the maximum accretion rate for stable H-shell burning on the surface of the WD, the unprocessed material is expelled in the form of an optically thick wind (see, e.g., \citealt{wang18}). If the WD explodes during the wind phase, the resulting SN~Ia could show the presence of hydrogen \citep{hachisu99,kato12}. 
Alternatively, the high mass transfer rate may lead to a long series of relatively mild nova eruptions. These nova cycles cause the initial CO WD to grow in mass until it reaches masses of the order of 1.35 - 1.38 M$_{\sun}$ and finally to explode as SN (see, e.g., \citealt{hillman16} or \citealt{hachisu18}). This is not the first time that a recurrent nova is proposed as the progenitor of SNe Ia, although this scenario was more often called to explain SNe strongly interacting with the CSM art early times (see, e.g., \citealt{dilday12,woodvasey06,hachisu18}).\\

There may be other possibilities that we have not discussed in detail here. For example, as recalled by \citet{dessart20}, the enclosed ($>$0.1 M$_{\sun}$) H-rich material that moves at low velocity may originate from a tertiary star in a triple system or a swept-up giant planet rather than from a non-degenerate companion star \citep{soker19}. \\

Altogether, the observables of SNe~2016jae, 2018cqj, and 2018fhw suggest that these could be transition objects between ``normal'' SNe~Ia and low peak luminosity SNe~Ia. The fact that these three supernovae share among the many similarities the presence of H at similar dates is suggestive and raises the possibility of a common progenitor system. 
Certainly, {\it the Zorro Diagram}, which includes a large sample of well-observed luminous, ``normal'' and sub-luminous SNe~Ia \citep{mazzali07}, suggests that all the SNe considered in this analysis have a similar mass progenitor, consistent with the Chandrasekhar model. However, \citet{dessart20} show a better fit of the optical spectrum of SN~2018cqj at 207d after explosion \citep{prieto20} with a mass that is lower than the Chandrasekhar mass. Indeed, there is growing evidence that the properties of low-luminosity supernovae are better explained by sub-Chandrasekhar mass explosion \citep{stritzinger06,scalzo19}. At the same time, from a theoretical point of view, it has been ruled out the single degenerate scenario for transitional SNe~Ia based on the t$_0$-M$_{^{56}Ni}$ relation \citep{wygoda19,sharon20}. 
However, the appearance of hydrogen at nebular phases appears to be not generally expected from the double degenerate scenario and in turn with the sub-Chandrasekhar model for the three SNe of our sample. In short, at the present stage, we cannot link the ``coincidence'' of these three SNe to a unique evolution scenario or explosion mechanism. 


%

\begin{acknowledgements}
      N.E.R. thanks D. Kushnir, M. Hernanz and J. Isern for useful discussions and to the telescope's staff for their excellent support in the execution of the observations.\\
      N.E.R. and E.C. acknowledges support from MIUR, PRIN 2017 (grant 20179ZF5KS). 
      Support for J.L.P. is provided in part by ANID through the Fondecyt regular grant 1191038 and through the Millennium Science Initiative grant ICN12$\_$009, awarded to The Millennium Institute of Astrophysics, MAS.      
      Based on observations made with the GTC telescope, in the Spanish Observatorio del Roque de los Muchachos of the Instituto de Astrofísica de Canarias, in the island of La Palma.
      This paper includes data gathered with the 6.5 meter Magellan Telescopes located at Las Campanas Observatory, Chile.
      We acknowledge ESA Gaia, DPAC and the Photometric Science Alerts Team (http://gsaweb.ast.cam.ac.uk/alerts).
      This work is based (in part) on observations collected at the European Organisation for Astronomical Research in the Southern Hemisphere, Chile as part of PESSTO, (the Public ESO Spectroscopic Survey for Transient Objects Survey) ESO program 188.D-3003, 191.D-0935, 197.D-1075.
      This research has made use of the NASA/IPAC Extragalactic Database, which is funded by the National Aeronautics and Space Administration and operated by the California Institute of Technology.
      
\end{acknowledgements}

%
%

\bibliographystyle{aa} 
\bibliography{sne1a} 

%

\begin{appendix} 
%
\section{Additional figures of SN~2016jae.}

\begin{figure}[!ht]
\centering
\includegraphics[width=\columnwidth]{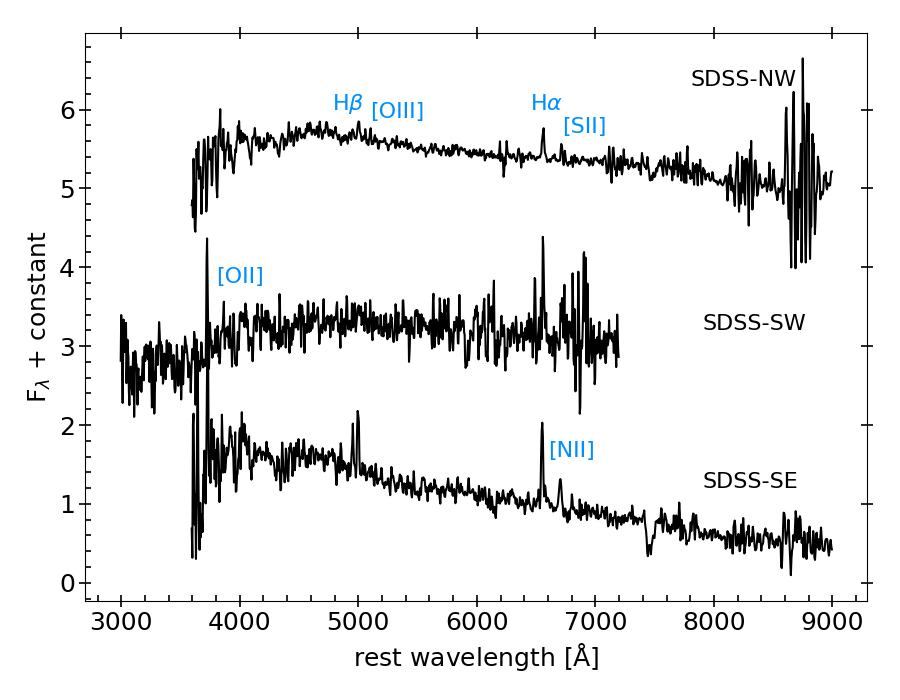} 
\caption{Spectra of SDSS-NW, SDSS-SW and SDSS-SE taken with GTC+OSIRIS on 2019 December 3, and 2020 May 10. In particular, the spectrum of SDSS-SW is a combination of the spectra of both epochs. All spectra are in the rest frame and corrected for Galactic extinction.} 
\label{fig_spechost}%
\end{figure}

\begin{figure}[!ht]
\centering
\includegraphics[width=\columnwidth]{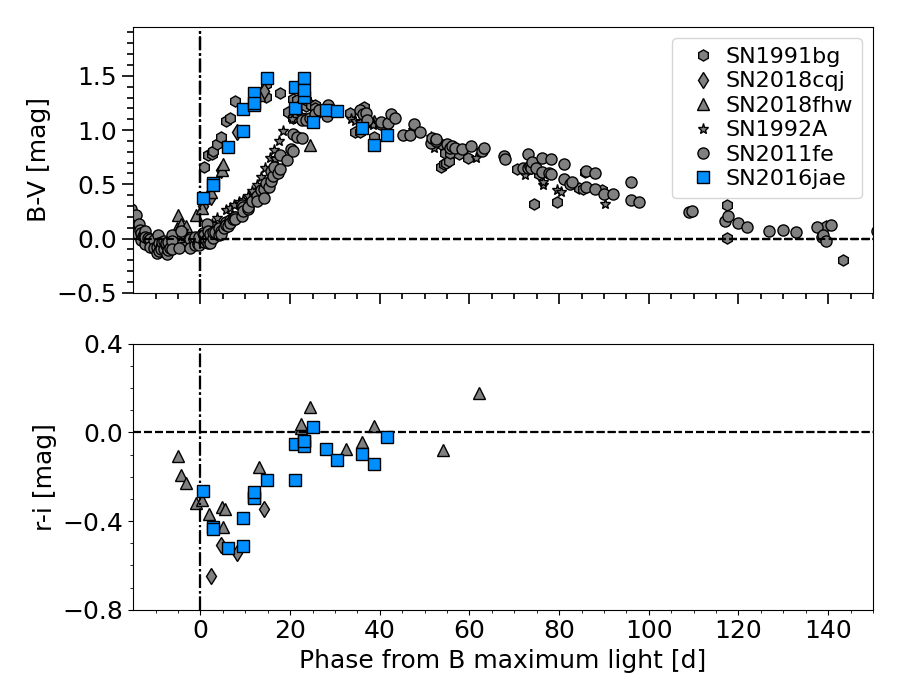} 
\caption{Intrinsic colour evolution of SN~2016jae, compared with those of SNe~1991bg, 2018cqj, 2018fhw, 1992A and 2011fe. For both panels, the dot-dashed vertical line indicates the $B$-band maximum light. 
} 
\label{fig_colourevol}%
\end{figure}

\begin{figure}[!ht]
\centering
\includegraphics[width=\columnwidth]{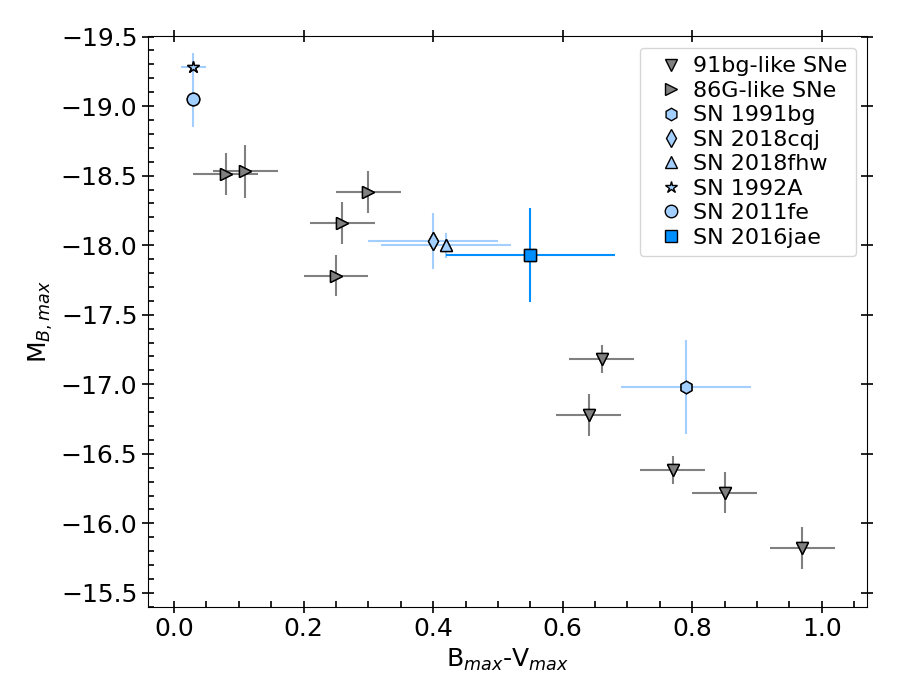} 
\caption{(B$_{max}$-V$_{max}$) colour vs. M$_{B_{max}}$ of a sample of sub-luminous SNe~Ia (91bg-like) and transitional or 86G-like SNe from the Carnegie Supernova Project-I sample of \citep{krisciunas17}. The magnitudes were derived using SNooPy light curve fits and the absolute magnitudes were calculated assuming H$_0$ = 67.8 km\,s$^{-1}$\,Mpc. Magnitudes of SNe~1991bg, 2018cqj, 2018fhw, 1992A, 2011fe and 2016jae are also plotted.} 
\label{fig_BVvsMB}%
\end{figure}

\begin{figure}[!ht]
\centering
\includegraphics[width=\columnwidth]{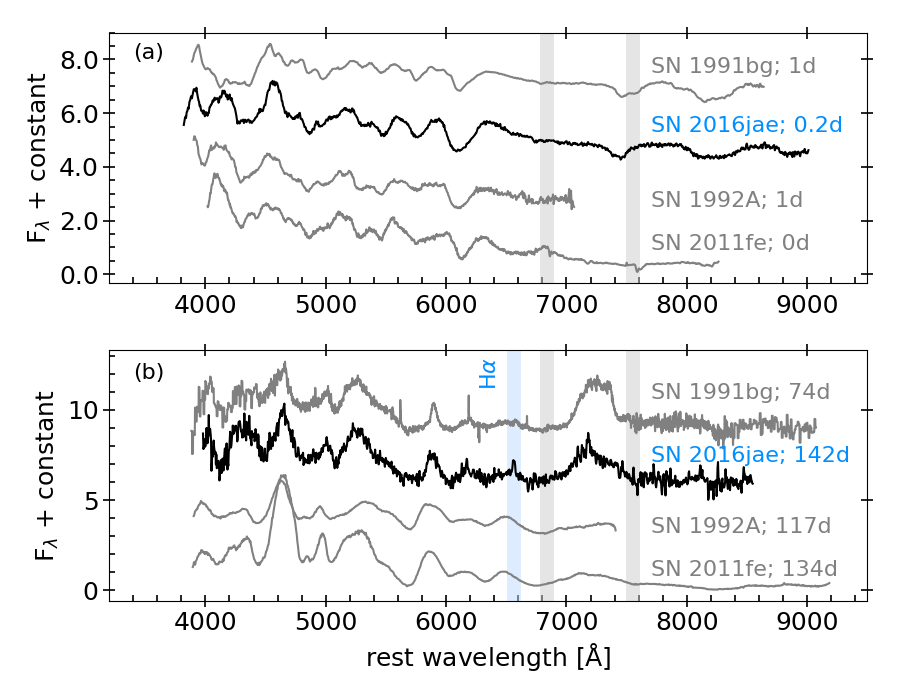} 
\caption{Early- ({\it (a)}) and late- ({\it (b)}) time comparison of SN~2016jae optical spectra, along with those of the sub-luminous SN~1991bg and the ``normal'' Ia~SNe~1992A and 2011fe at similar epochs. All spectra have been corrected by redshift and extinction (adopted values and references are reported in Table \ref{table_SNe}). Ages are relative to $B$ maximum light. The grey columns show the location of the strongest telluric band, which has been removed when possible.}
\label{fig_spececomp_earlylate}%
\end{figure}

\begin{figure}[!ht]
\centering
\includegraphics[width=\columnwidth]{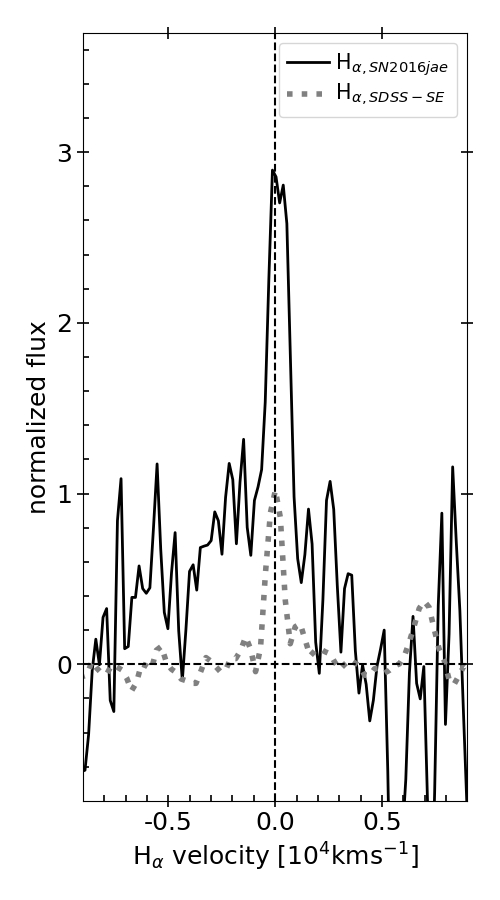} 
\caption{H$\alpha$ line profile comparison of SN~2016jae at phase 141.8 d (from the assumed $B$ maximum date) and the spectrum of SDSS-SE (the nearest galaxy with a similar redshift). The dashed lines mark the rest wavelength of H$\alpha$ and the normalized flux = 0.}
\label{fig_halphacomp_SNhost}%
\end{figure}

%
\section{Properties of the supernovae used in this work.}

\begin{table*}[!ht]
 \centering
  \setlength\tabcolsep{2.5pt}
  \caption{Properties of the supernovae used along this work.}
  \label{table_SNe}
  \begin{tabular}{@{}llcccc@{}}
  \hline
 SN   & Host galaxy & Distance$^{\dagger}$ & $E(B-V)_{\rm tot}$  &   $\sim$ t$_{B,max}$ & Sources  \\
        &  & (Mpc) & (mag) & (MJD) & \\
 \hline
SN~1991bg & NGC~4374 & 19.1 & 0.08 & 48603.2 & a \\
SN~1992A & NGC~1380 & 22.3 & 0.01 & 48640.0 & b \\
SN~2011fe & M101     & 6.2  & 0.02 & 55815.5 & c \\
SN~2018cqj & IC550   & 78.9 & 0.03 & 58294.7 &  d \\
SN~2018fhw & 2MASX J04180598–6336523 & 76.2 & 0.03 & 58356.8 & e \\
SN~2016jae & ?       & 92.9 & 0.10 &  57750.2 & This work \\
\hline
\end{tabular}
\begin{flushleft}
$^{\dagger}$ Distances have been scaled to $H_0$ = 67.8 \kms Mpc$^{-1}$.\\
a=\citealt{filippenko92,leibundgut93,turatto96}; 
b=\citealt{phillips99,stritzinger10};
c=\citealt{pereira13,silverman12,maguire14}; 
d=\citealt{prieto20}; 
e=\citealt{kollmeier19,vallely19}. 
\end{flushleft}
\end{table*}

%
\section{Tables of photometry and spectroscopy of SN~2016jae.}

\begin{table*}
\centering
\caption{New $r$ ({\sc ABmag}) photometry of SN~2016jae.}
\label{table_ph}
\begin{tabular}{@{}ccccc@{}}
\hline 
Date & MJD & Phase$^a$  & r & Instrument key$^b$ \\ 
 &  & (days) & (mag) &  \\ 
\hline 
20161222 & 57744.28 &    -5.9 &  17.39 (0.17) & Gaia  \\ 
20161230 & 57752.36 &     2.2 &   16.95 (0.15) & Gaia  \\ 
20170330 & 57842.98 &    92.8 & 20.85 (0.19) & Gaia  \\ 
20170518 & 57891.91 &   141.8 & 22.53 (0.09) & OSIRIS  \\ 
\hline  
\end{tabular}
\begin{flushleft}
$^a$ Phases are relative to $B$ maximum light, MJD = 57750.15 $\pm$ 1.00.\\ 
$^b$ Gaia; OSIRIS = 10.4 m Gran Telescopio CANARIAS + OSIRIS located at the Roque de Los Muchachos, Spain\\
\end{flushleft}
\end{table*}

\begin{table*}
 \centering
  \caption{Log of spectroscopy observations of SN~2016jae.}
  \label{table_spec}
\scalebox{0.9}{
  \begin{tabular}{@{}ccccccccc@{}}
  \hline
  Date & MJD & Phase$^a$ & Instrumental set-up$^b$ & Grism or grating & Spectral range & Resolution$^c$ & Seeing & Airmass\\
 & & & & & (\AA) & (\AA) & (arcsec) & \\
 \hline
20161228 & 57750.34 & 0.2   & NTT+EFOSC2 & gm13+1.00 arcsec & 3900-9230 & 18 & 1.6 & 1.32\\
20170322 & 57834.19 & 84.0  & Clay/Magellan+LDSS-3 & VPH-All+1.00 arcsec & 3800-9285 & 8 & 1.1 & 1.67 \\
20170518 & 57891.91 & 141.8 &  GTC+OSIRIS & R500B+1.00 arcsec & 4068-8730 & 7 & 1.4 & 1.44\\  
\hline
\end{tabular}
}
\begin{flushleft}
$^a$ Phases are relative to $B$ maximum light, MJD = 57750.15 $\pm$ 1.00.\\ 
$^b$ NTT+EFOSC2 = 3.6m New Technology Telescope + EFOSC2 located at the La Silla Observatory, Chile; Clay/Magellan+LDSS-3 = 6.5m Magellan Clay Telescope + LDSS3 (Low Dispersion Survey Spectrograph 3) located at Las Campanas Observatory, Chile; GTC+OSIRIS = 10.4 m Gran Telescopio CANARIAS + OSIRIS losted at the Roque de Los Muchachos, Spain.\\
$^c$ Measured from the FWHM of the night sky lines.
\end{flushleft}
\end{table*}

\end{appendix}
\end{document}